\documentclass[12pt,trackchanges,preprint]{aastex63}
\usepackage{amssymb,amsbsy,amsmath}
\usepackage{verbatim,palatino,placeins}
\usepackage{graphicx}
\begin{document}

\title{Validation \& Testing of the CROBAR 3D Coronal Reconstruction Method with a MURaM simulation}
\correspondingauthor{Joseph Plowman}
\email{jplowman@boulder.swri.edu}

\author[0000-0001-7016-7226]{Joseph Plowman}
\affil{Southwest Research Institute,
Boulder, CO 80302 USA}


\begin{abstract}
I report on validation and testing of a novel 3D reconstruction method than can obtain coronal plasma properties from a single snapshot perspective. I first reported on the method in 2021, and I have since named it the Coronal Reconstruction Onto $\mathbf{B}$-Aligned Regions, or `CROBAR', method. The testing was carried out with a cube from a MURaM 3D MHD simulation, which affords a coronal-like `ground truth' against which the reconstruction method can be applied and compared. I find that the method does quite well, recovering the `coronal veil'-like features recently reported from the MURaM simulations, and allaying concerns that these features would thwart recovery of valid 3D coronal structure from a limited number of perspectives. I also find that a second perspective at between $\sim 45$ and 90 degrees, does significantly improve the reconstructions. Two distinct channels with Soft X-Ray like temperature response (peaking above 5 MK) would suffice for CROBAR's optically thin observables; barring that, a suite of AIA-like EUV passbands, with good coverage of the 3-8 MK temperature range.
\end{abstract}

\section{Introduction}


In \cite{Plowman_3DR2021}, I reported on a new method that can reconstruct 3D coronal plasma properties even from a single perspective. The method works by fitting emission profiles, as functions of field line arc length, to a set of volume-filling field aligned regions. As a result, the method has been named `Coronal Reconstruction onto B Aligned Regions', or CROBAR. 

The initial paper described the method and showed an initial application to some data from the Solar Dynamics Observatory Atmospheric Imaging Assembly \citep[SDO/AIA][]{sdo_paper, sdo_aia_paper}, as a proof of concept. The reconstructions shown there used only the AIA perspective for the reconstruction, and a potential field for the field-aligned structure. A subsequent comparison with observations from the Solar Terrestrial Relations Observatory \cite[STEREO][]{STEREO_Kaiser2008} sufficed to show the plausibility of the technique, albeit hampered by the limits of the potential field. 

In this paper, I show a validation of the field-aligned region reconstruction approach using a 3D MHD simulation from a MPS/University of Chicago Radiative MHD \citep[MURaM, e.g.,][]{Rempel_ApJ2017, CheungEtalNatAs2019} simulation as a synthetic `ground truth'. A recent paper using this simulation \citep{MalanushenkoEtal_coronalveil2022} has pointed out some shortcomings of the traditional coronal loops approach when it comes to recovering 3D coronal plasma structure using a limited number of structures. CROBAR has some parallels and grows out of a coronal loop-like approach, so it may be reasonable to wonder if it is not subject to these limitations. However, I believe CROBAR largely avoids the concerns expressed in \cite{MalanushenkoEtal_coronalveil2022}, for the following reasons:

\begin{enumerate}
	\item CROBAR is not a `blind' inversion. The magnetic field provides the `skeleton' which is fleshed out by the field-aligned regions. The fundamental physical assumption it relies on, which \cite{MalanushenkoEtal_coronalveil2022} does not invalidate, is that the plasma in the corona is mostly forced to follow the coronal magnetic field, and the field structure is largely determined by a boundary value problem. The reconstruction problem taking this into account is therefore largely a 2-Dimensional one, solvable with 2D data.
	\item The CROBAR B-aligned regions fill the entire volume, rather than being just a handful of loops on top of a static (or empty) background. There are also a large number of them. which enables them to fit 
	\item Cross-sectional shape the of regions follows and deforms as a function of arc length, again following the field structure. This can allow CROBAR to replicate the `coronal veil' appearance described by \cite{MalanushenkoEtal_coronalveil2022}
\end{enumerate}

Indeed, some of the same concerns raised in \citep{MalanushenkoEtal_coronalveil2022} motivated the development of these features in what ultimately become CROBAR. In this paper I focus on how well CROBAR performs at recovery of volume emission information when given a magnetic field and one or more optically thin emission images. That is, I use the magnetic field in the MURaM simulation to compute the skeleton on which the field aligned regions are built. The intention is to isolate each component of the validation (in this case the emission structure) and verify correct function in isolation. I leave the other major component (refinement of the magnetic field extrapolation based on the residuals of the emission structure) to a future paper, even though I have already implemented a basic version of this component based on a linear force-free field. Covering both of these topics simultaneously would result in a long and unfocused paper.

I now give a brief discussion of the MURaM simulation.

\subsection{MURaM simulation and Technical Considerations}

I use the same coronal MURaM simulation also reported in \cite{CheungEtalNatAs2019, MalanushenkoEtal_coronalveil2022}, however I use a somewhat earlier time in the simulation (The time of the snapshot may be referenced by its unique index of 100000) during which the volume is less active. The CROBAR approach is less likely to be suited to times of strong plasma dynamics when the field-alignment of the plasma is less certain and the field is less likely to be force free. This time interval in the simulation still contains the complex structures pointed out by \cite{MalanushenkoEtal_coronalveil2022} and shows the sort of complexity associated with solar active regions. It is important to note that there some differences with a typical solar active region, however:

\begin{itemize}
	\item The region is significantly smaller than the long-lasting active regions most often the focus of solar physics research, at roughly $100$ by $50$ by $50$ Megameters, or fewer than 200 pixels across as viewed by SDO/AIA. Many active regions are several times this size in linear extent.
	\item The boundary conditions are periodic, so that the region behaves as if it has an identical clone of itself on each side rather than being relatively isolated.
	\item The simulation performs a number of approximations to deal with aspects that are to computationally demanding to incorporate (such as extremely high Alfven wave speeds in some regions), and assumes the chromosphere is LTE.
\end{itemize}

For more discussion of the details of the simulation, see \cite{Rempel_ApJ2017, CheungEtalNatAs2019, MalanushenkoEtal_coronalveil2022}. Figures showing the appearance of the simulation will be shown in connection with the reconstructions, as the narrative develops.

\section{CROBAR Methodology and MURaM Forward Modeling}
\subsection{Magnetic Field Skeleton and B-aligned regions}
As previously mentioned, this paper treats only the plasma emission part of the problem. The emphasis is on the question of whether field aligned emission can reproduce coronal emission, at all, regardless of the quality of the magnetic field skeleton. There is little point in covering the refinement of the magnetic field skeleton before this question has been addressed, as \cite{MalanushenkoEtal_coronalveil2022} has illustrated. Therefore, I use as the skeleton the magnetic field contained in the MURaM simulation rather than (for example) attempting to use and refine a linear force-free field. I intend to address that topic in a future paper, and have already carried out some work in that direction with promising results.

I traced 10,500 field lines through the volume, using the MURaM field vectors. The tracing was done with the 2/3 order Runge-Kutta scheme included in scipy's \texttt{integrate.solve\_ivp} method. 10,000 of those initial points were at the photosphere level and their location was randomly chosen with a weight based on the vertical field strength at the photospheric level. The other 500 initial points were uniformly distributed through the volume and then traced in both directions until they reached the photosphere or the edges of the cube. I removed field lines from the set which were too short in either height (less than one voxel, 0.064 Mm) or overall length (less than 8 voxels, 0.512 Mm). This is referenced to the voxel grid used for associating points in the volume with field lines, which was chosen to have 0.064 Mm grid size in each direction rather than the $192 \times 192 \times 64$ Mm grid resolution used in the MURaM simulation. This choice was made to ensure full resolution of the simulation and also because the computational load if CROBAR is light. The voxel grid used by CROBAR therefore has just under 1 billion voxels, requiring 3 GB of RAM (2 GB for the array indicating the loop ID of each voxel, and 1 GB for the array indexing its length along the loop).

\begin{figure}[!htbp]
	\begin{center}
		\includegraphics[width=0.8\textwidth]{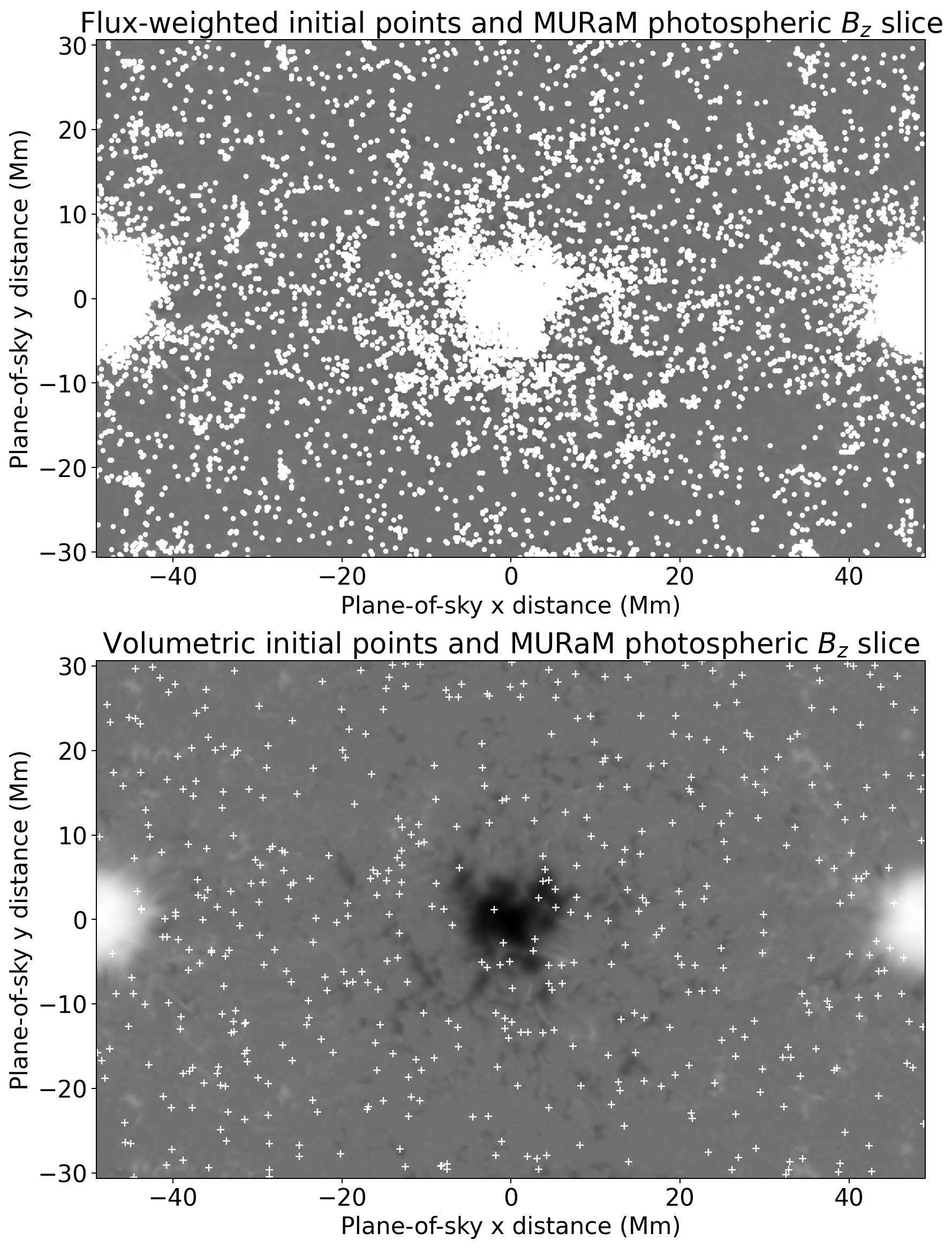}
	\end{center}
	\caption{Initial points for CROBAR fieldline tracing: Upper panel shows flux-weighted initial points, which start at the photospheric level in the MURaM simulation, lower panel shows the volume initial points which are uniformly weighted throughout the volume. Both are plotted over the MURaM vertical field strength at the photosphere, analogous to a longitudinal mangetogram.}\label{fig:photospheric_field_footpoints}
 \end{figure}

\begin{figure}[!htbp]
	\begin{center}
		\includegraphics[width=0.39\textwidth]{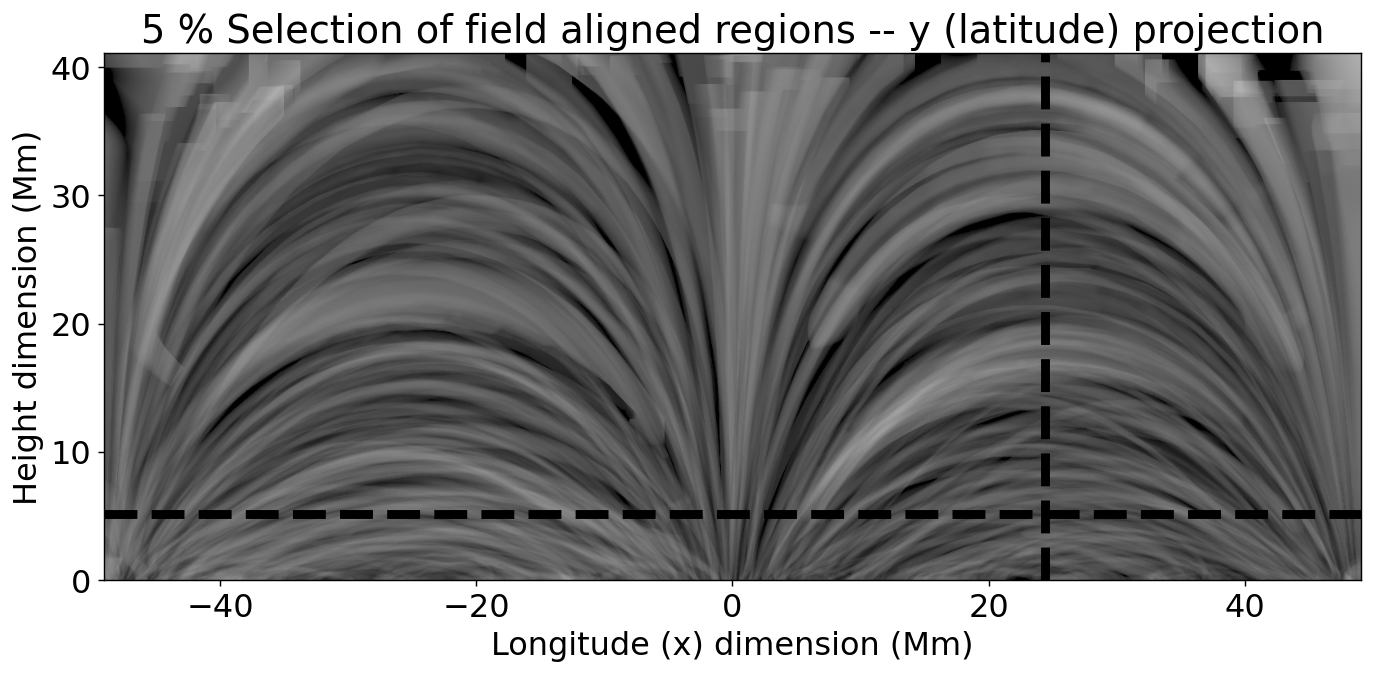}\includegraphics[height=0.2035\textheight]{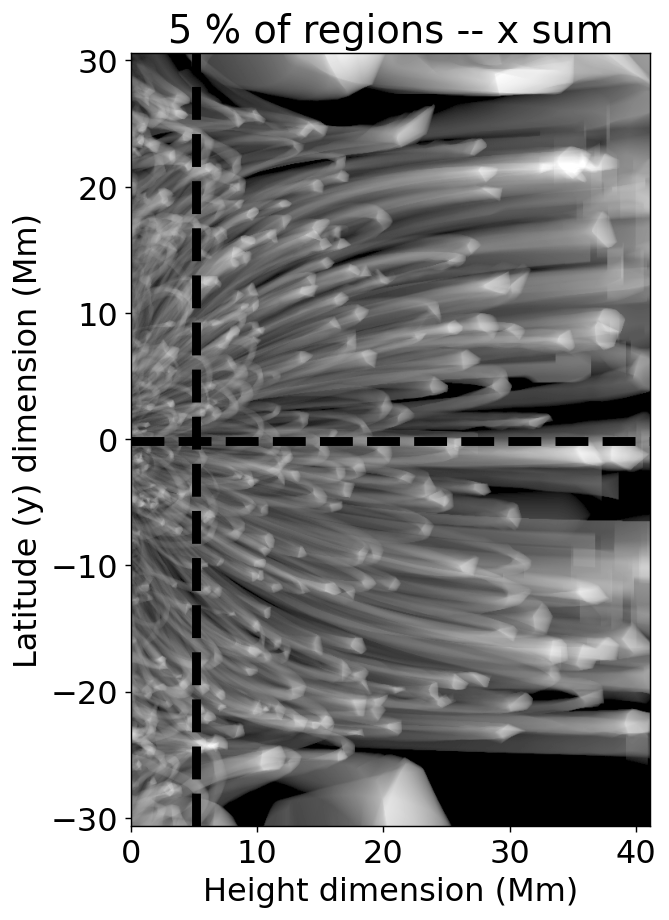}\includegraphics[height=0.2035\textheight]{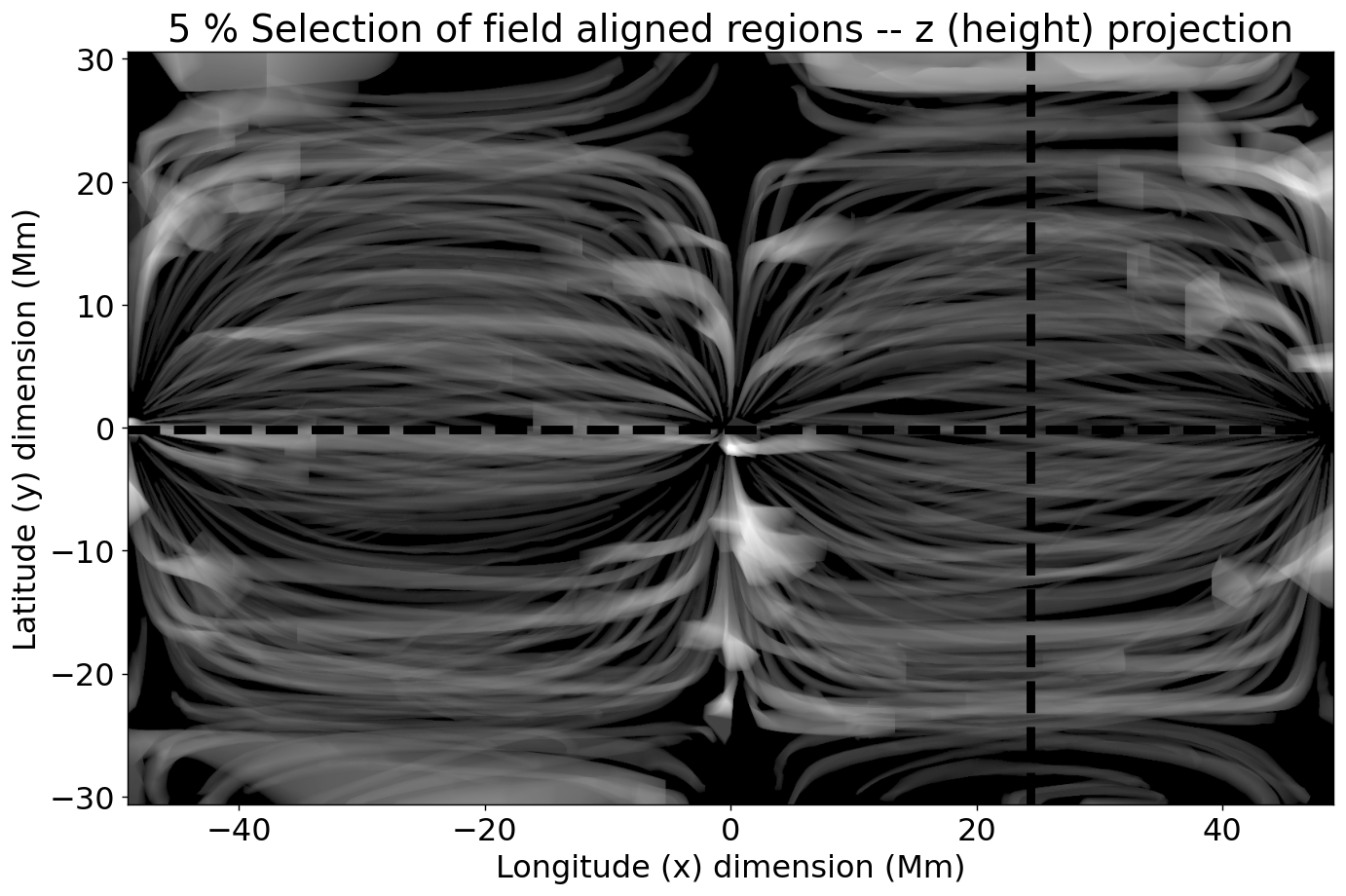}\\
		\includegraphics[height=0.2035\textheight]{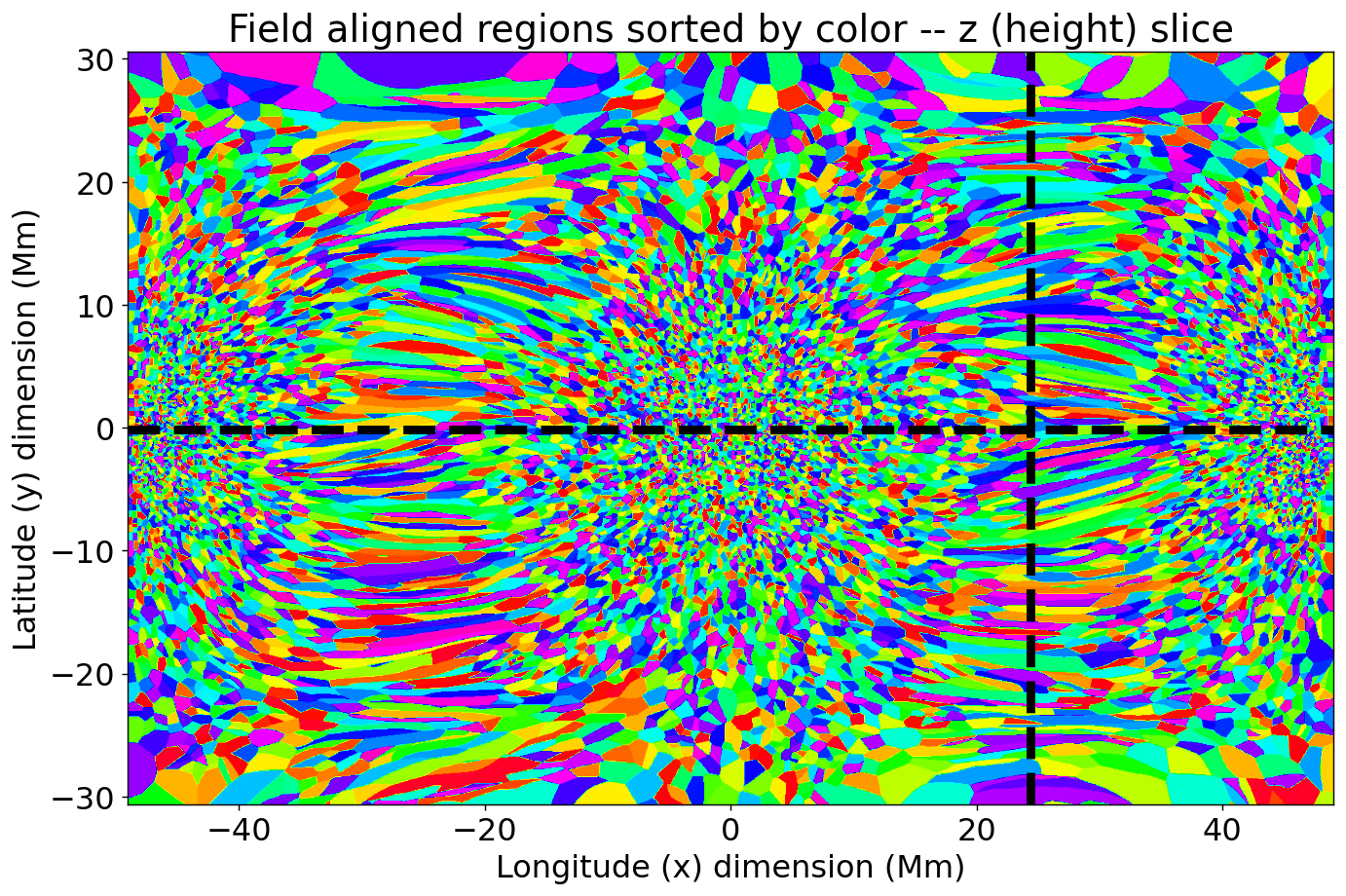}\includegraphics[height=0.2035\textheight]{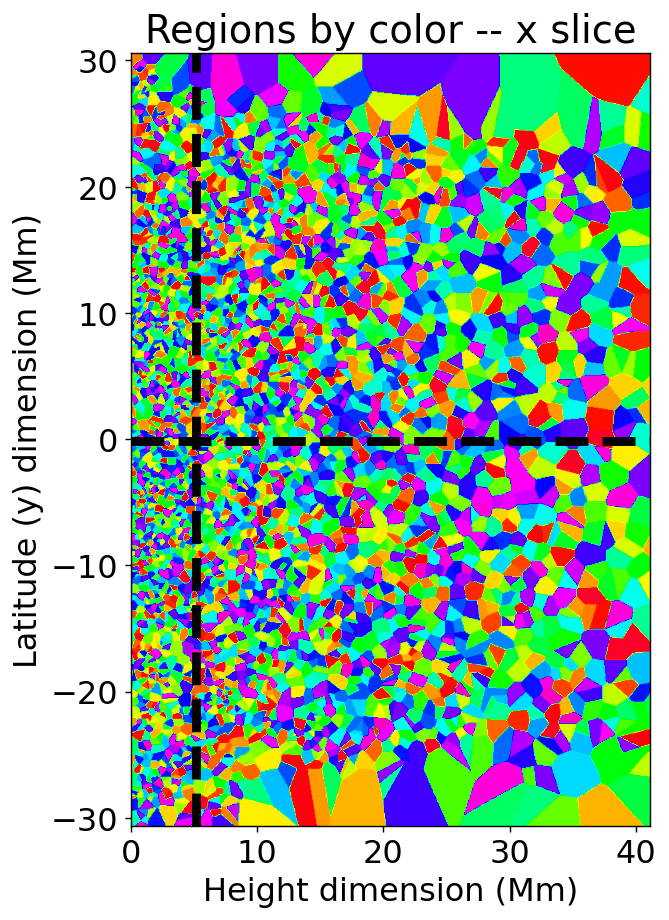}\includegraphics[width=0.39\textwidth]{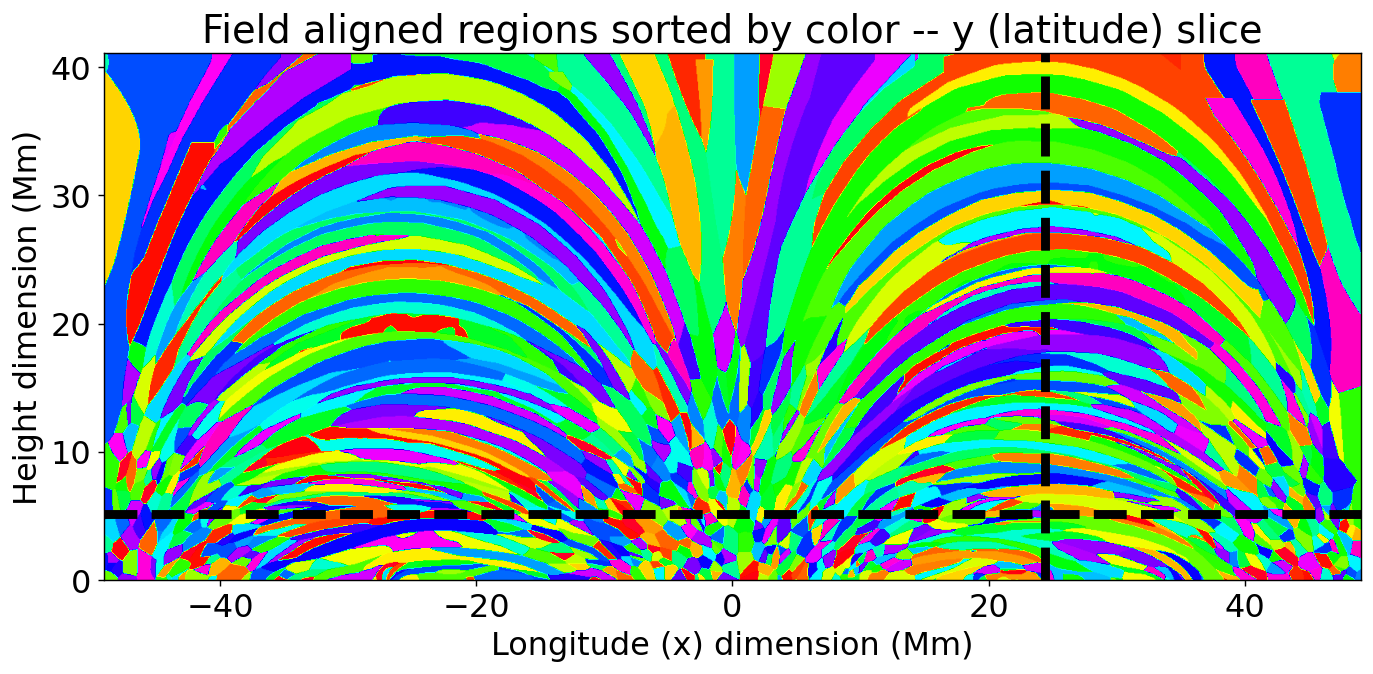}\\
	\end{center}
	\caption{Figure showing field-aligned regions used for the CROBAR reconstructions of the MURaM cube. Projections of a subset of the region are shown in the top row, while slices along the principal axes are shown on the lower row. Different colors in the lower plot represent different regions, and the dashed lines in each plot show the locations of the slices. The x slice will also be shown in plots of the MURaM emission and the reconstructions.}\label{fig:BAR_example_plots}
 \end{figure}

Images of the seed points for tracing fieldlines and the vertical field component at the nominal photospheric level \citep[8 Mm, 125 voxel heights, compare][]{Rempel_ApJ2017} are shown in Figure \ref{fig:photospheric_field_footpoints}. The boundary conditions are at least nominally circular, and I wanted to maximize the number of complete field lines (footpoint to footpoint), so I shifted the cube 128 voxels to the left compared to its usual presentation, placing the negative polarity at cube center. This makes nearly all loops in the cube complete, since most large-scale connections in the cube are between the positive polarity/ies and the negative polarity/ies. I also padded the cube on the top and bottom (using the circular boundary conditions) by 32 pixels, in a further effort to avoid edge effects. The field-aligned regions used for the reconstruction are shown in Figure \ref{fig:BAR_example_plots}, using both a projection through with the voxels of 1 in 20 of the regions set to one, as well as a set of orthogonal slices through salient locations in the cube.

The padded voxels mentioned above are omitted from the reconstruction, since they contain incomplete field lines. Similarly, I noticed a small edge discontinuity in the cube at 128 voxels, so I only used the right-hand-side (rightmost 256 voxels) of the cube for the reconstructions; emission from voxels outside the resulting $49.152 \times 49.152 \times 41.152$ Mm subsection of the MURaM cube is set to zero in both the forward modeling and the CROBAR reconstructions. Emission from outside this subsection will not be shown in subsequent figures.

\subsection{Optical Depth and Related Considerations}\label{sec:depth_related_considerations}

One part of the forward and inverse modeling that is essential to consider is the transition from optically thin emission to optically thick in whatever part of the solar spectrum is being observed. Although coronal emission is generally described as optically thin, at some point it transitions to being optically thick; otherwise we could see through the solar disk in these passbands! 


This is complicated by the simplifying assumptions made in the MURaM simulations. While necessary for computational tractability, they result in high-density cusps at the base of the observing region which can appear to dominate the emission, unlike real solar passband images. CROBAR does not accurately reproduce these, which leads to issues with the reconstructions. To avoid these issues, I set a minimum height cutoff of 2.5 Megameters above what I take to be the photospheric height, which in turn is 8 Megameters (125 voxel heights) above the base of the simulation. Above this height, I approximate optical depth attenuation with a local extinction factor $e^{-\tau(z)}$ \citep[compare][]{Mok2016} where $\tau(z)$ which is proportional to the vertical integral of the density. Naturally, this is inexact for non-vertical lines of sight. The constant of proportionality is set so that optical depth unity occurs at a column density of $3\times 10^{19} \mathrm{cm}^{-2}$.

These issues do also suggest that CROBAR is not as readily suited to passbands or lines which have a significant contribution from non-optically thin emission at the footpoints, although I expect such difficulties can be mitigated with further refinement and iteration. This is not simply a question of where the temperature response function is large, however; because emission is proportional to temperature response times density squared ($R(T)n^2$), one must also consider the density where optically thick emission becomes significant. The condition results in the following inequality:
\begin{equation}
	R(T(\tau=1)) < R(T(\tau << 1)) \big[\frac{n(\tau=1)}{n(\tau <<1)}\big]^2
\end{equation}
if the emission from optically thin temperatures $T(\tau << 1)$ is to dominate over the optically thick emission $T(\tau = 1)$. If the ideal gas law holds and the pressure scale height is large, this becomes
\begin{equation}
	R(T(\tau=1)) < R(T(\tau << 1)) \big[\frac{T(\tau << 1)}{T(\tau=1)}\big]^2.
\end{equation}
In other words, the $R(T)$ at optically thin temperatures must be greater than at $\tau=1$ temperature by a factor of the {\em square} temperature ratio $T(\tau << 1)/T(\tau=1)$. This may have underappreciated implications for the temperatures of other features observed in the corona such as the moss; The temperature of the most visually apparent features at low heights may be much cooler than the shape of the temperature response function alone would indicate.

\subsection{Forward Modeling of the MURaM cube}

\begin{figure}[!htbp]
	\includegraphics[height=0.375\textwidth]{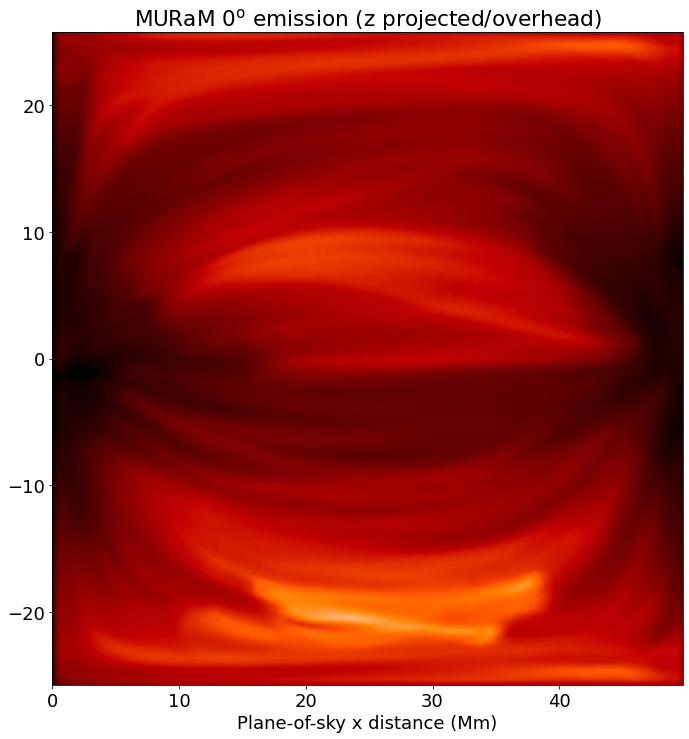}\includegraphics[height=0.375\textwidth]{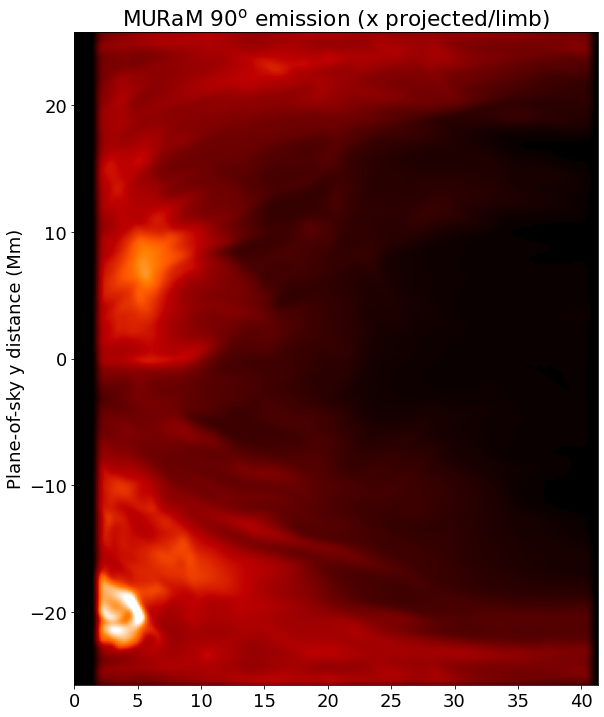}\includegraphics[height=0.375\textwidth]{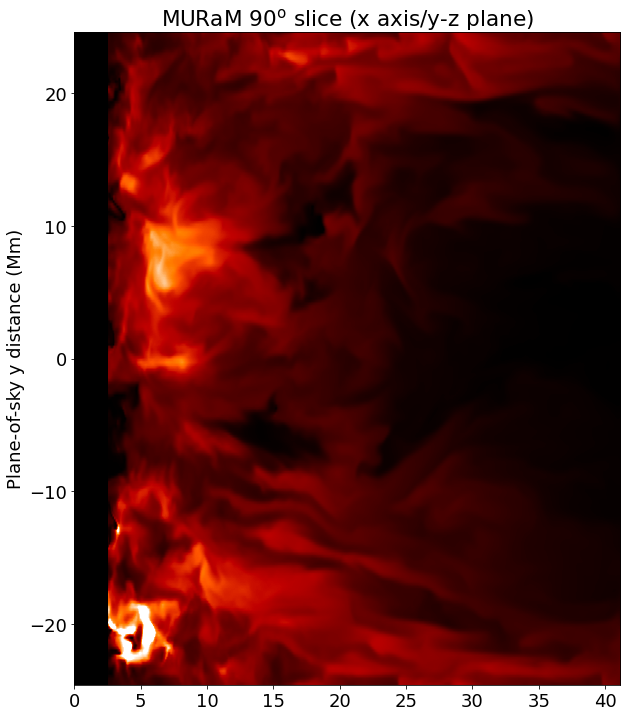}\\
	\includegraphics[height=0.375\textwidth]{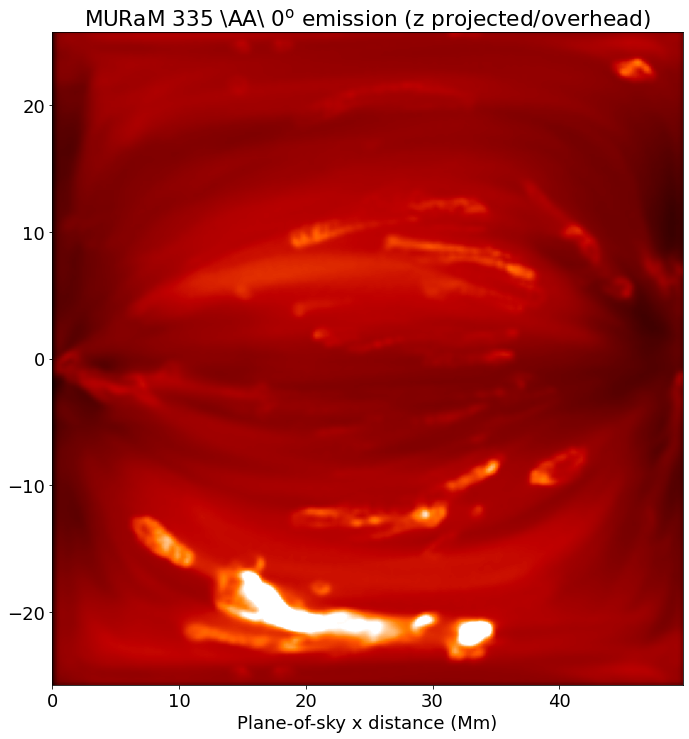}\includegraphics[height=0.375\textwidth]{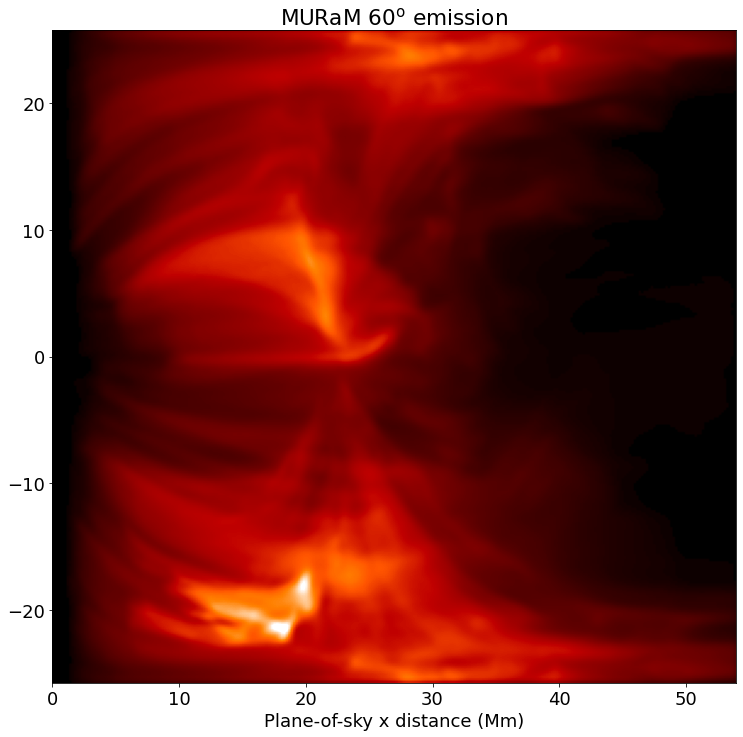}\\
	\caption{Emission for the MURaM data cube. Top left is the overhead ($0^\circ$) forward modeled emission using the optimal (for CROBAR) $T^2$ response function, lower left shows AIA 335 \AA\ for comparison (bright `cuspy' features in 335 are likely due to simplifying assumptions in MURaM; see discussion in text). Top center shows limb ($90^\circ$) forward modeled emission, while lower center shows $60^\circ$ forward modeled emission.}\label{fig:MURaM_fwd_synth_projections}
 \end{figure}

The temperature response function for this exercise was the power with law index $2$ mentioned in \cite{Plowman_3DR2021}. This is the ideal response function for this purpose because the emission profile of each region is a function only of pressure and dependence on specific details of the region's temperature profiles is minimized. The linearity assumption is therefore most nearly satisfied. As \cite{Plowman_3DR2021} noted, this kind of response function can be synthesized from a DEM, and it is also very similar to some lower X-Ray response functions such as from the Hinode X-Ray Telescope \citep[XRT][]{HinodeXRT} or Yohkoh Soft X-Ray Telescope \cite[SXT][]{YohkohSXT}. I have also noticed that they tend look quite similar to images from AIA 335 \AA , suggesting that a large fraction of coronal emission comes from temperatures where that passband has a scaled powerlaw index of 2. This is a potentially interesting result, which I intend to investigate further. For the reconstructions in this paper, I will consider viewpoints of $0^\circ$ (overhead) alone, $0^\circ + 60^\circ$, and $0^\circ + 90^\circ$ ($90^\circ$ corresponds to a typical limb view of the region). those are each shown in Figure \ref{fig:MURaM_fwd_synth_projections} along with an equivalent overhead view with the AIA 335 \AA\ temperature response function (The latter shows more `cuspy' low-lying emission than the $T^2$ channel, more so than is typical with real AIA 335 \AA\ data, so I expect those features are due at least in part to the simplifying assumptions of MURaM previously mentioned). I also show a longitudinal slice through the region, illustrating that this emission contains the interesting structures noted by \cite{MalanushenkoEtal_coronalveil2022}. Each of these views of the MURaM cube will also be shown alongside the reconstructions.

\section{CROBAR Results}

For the CROBAR profiles I assumed an RTV-like profile with an apex temperature of 2.5 MK and an exponential pressure profile with a scale height of 60 Mm/MK (150 Mm for 2.5 MK apex temperature), although because I use the $T^2$ power law for the temperature response only the pressure profile affects the results. I also allow the loops to be asymmetric in their emission profile, with separate coefficients controlling the left and right sides of the loop which are averaged at the midpoint. Field-aligned regions can have very different geometry at their endpoints, and I prefer to let the observations dictate this behavior rather than imposing symmetry.

CROBAR is built around a $\chi^2$ minimization, which requires a specification of measurement errors or equivalent. In this case the measurement errors are typically the shot noises in the EUV or X-Ray observations, so I assume that the errors are proportional to the square root of the forward modeled intensities, with the constant of proportionality set so that the median error level in the overhead data is $5\%$. I also added a small amount of $l^2$ norm regularization, enough to smooth the results but not enough to have a significant effect on the goodness of fit.

\begin{figure}[!htbp]
	\includegraphics[width=0.5\textwidth]{MURaM_emission_0.png}\includegraphics[width=0.5\textwidth]{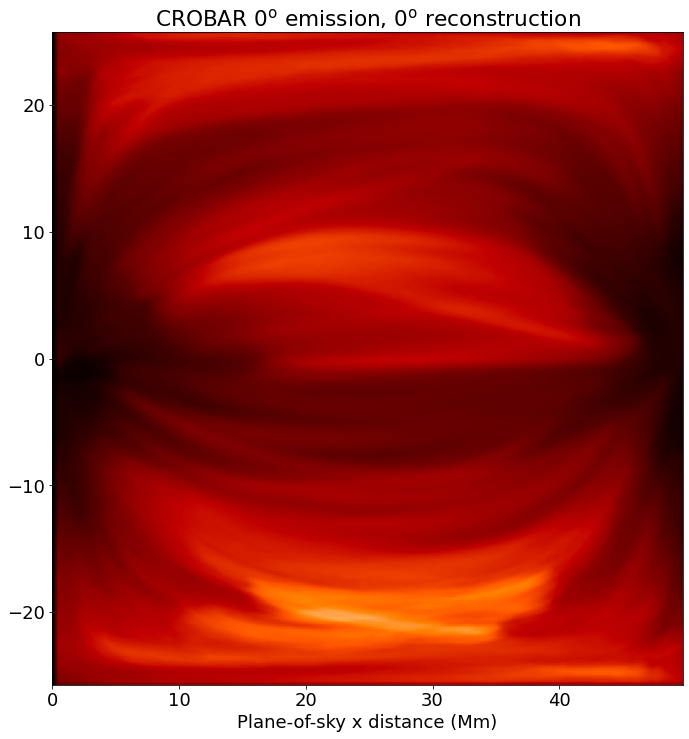}\\
	\includegraphics[width=0.5\textwidth]{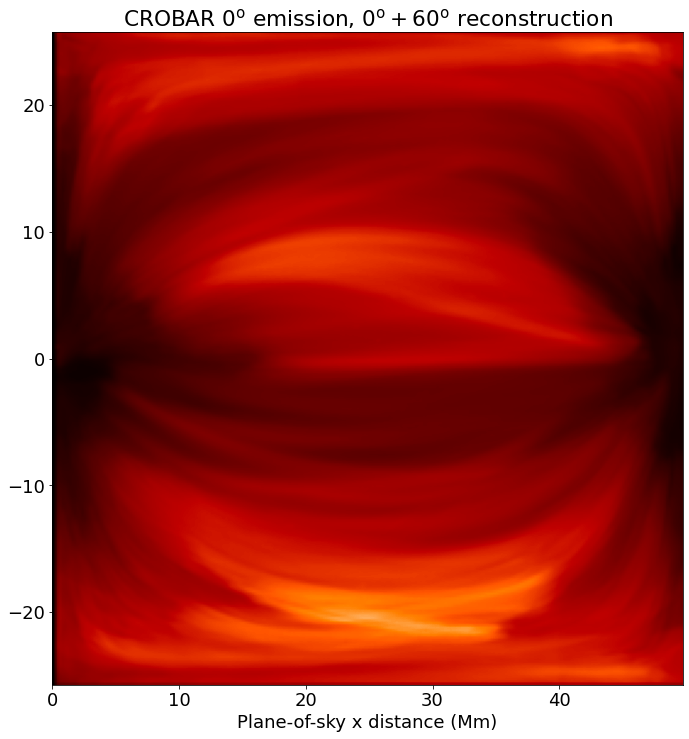}\includegraphics[width=0.5\textwidth]{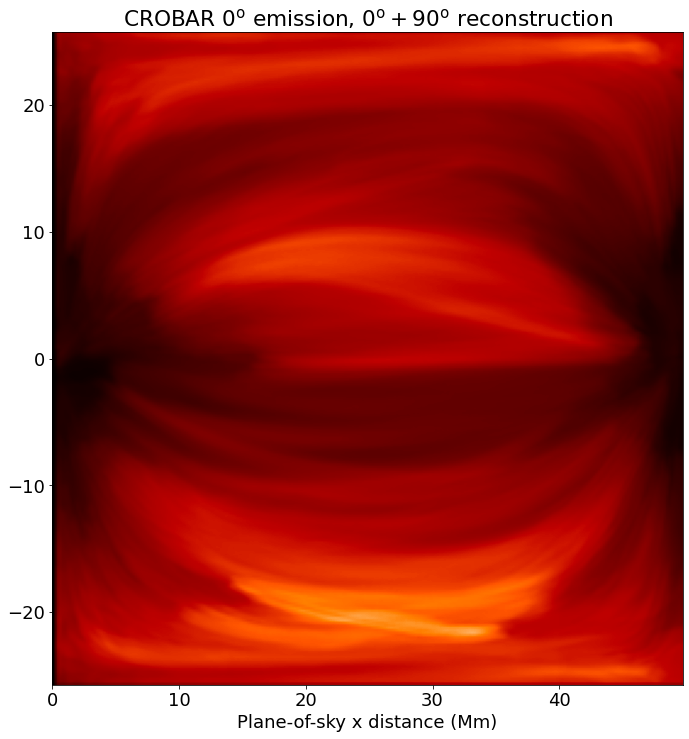}\\
	\caption{CROBAR reconstruction of the MURaM 0 degree emission. Top left shows original MURaM emission, top right shows CROBAR reconstruction using only the MURaM 0 degree emission, lower left shows CROBAR reconstruction using MURaM 0 and 90 degree emission, lower right shows CROBAR reconstruction using the MURaM 0 and 60 degree emission.}\label{fig:reconstructions_overhead}
 \end{figure}

\begin{figure}[!htbp]
	\includegraphics[width=0.5\textwidth]{MURaM_emission_60.png}\includegraphics[width=0.5\textwidth]{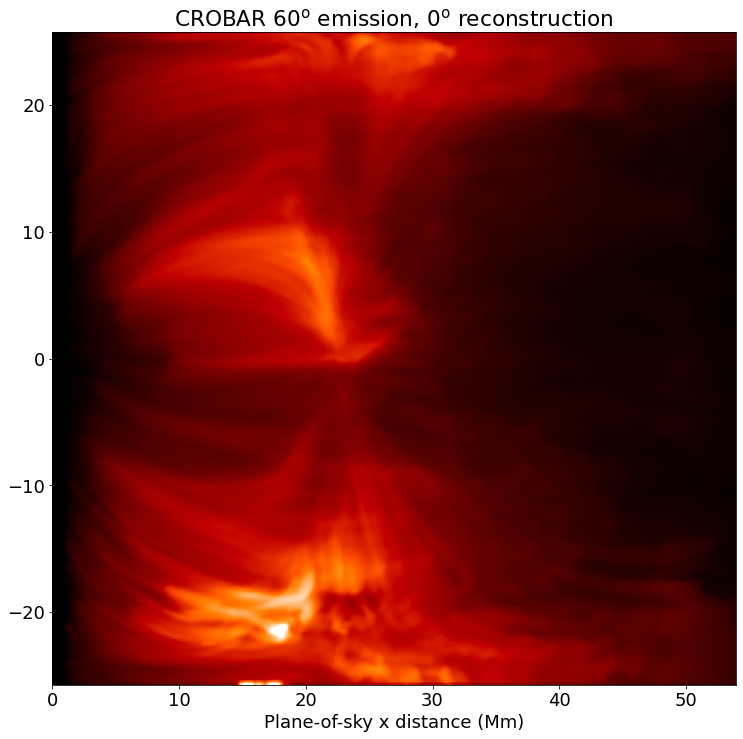}\\
	\includegraphics[width=0.5\textwidth]{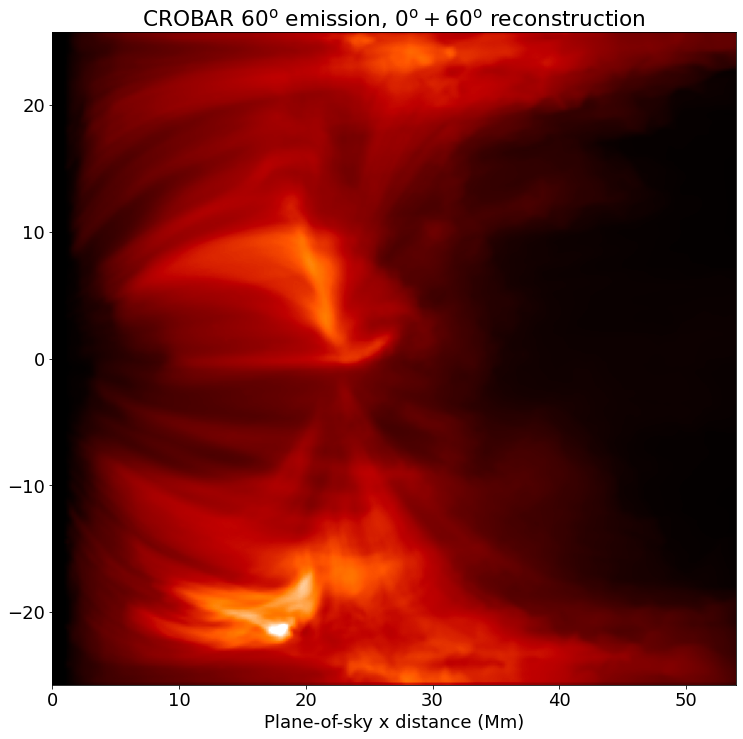}\includegraphics[width=0.5\textwidth]{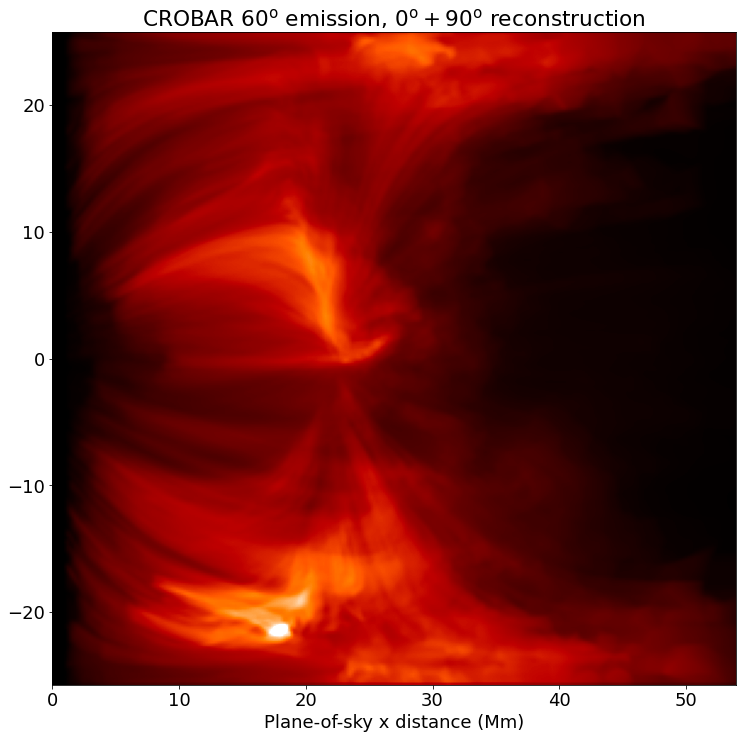}\\
	\caption{CROBAR reconstruction of the MURaM 60 degree emission. Top left shows original MURaM emission, top right shows CROBAR reconstruction using only the MURaM 0 degree emission, lower left shows CROBAR reconstruction using MURaM 0 and 90 degree emission, lower right shows CROBAR reconstruction using the MURaM 0 and 60 degree emission.}\label{fig:reconstructions_60}
 \end{figure}

\begin{figure}[!htbp]
	\includegraphics[width=0.5\textwidth]{MURaM_emission_90.png}\includegraphics[width=0.5\textwidth]{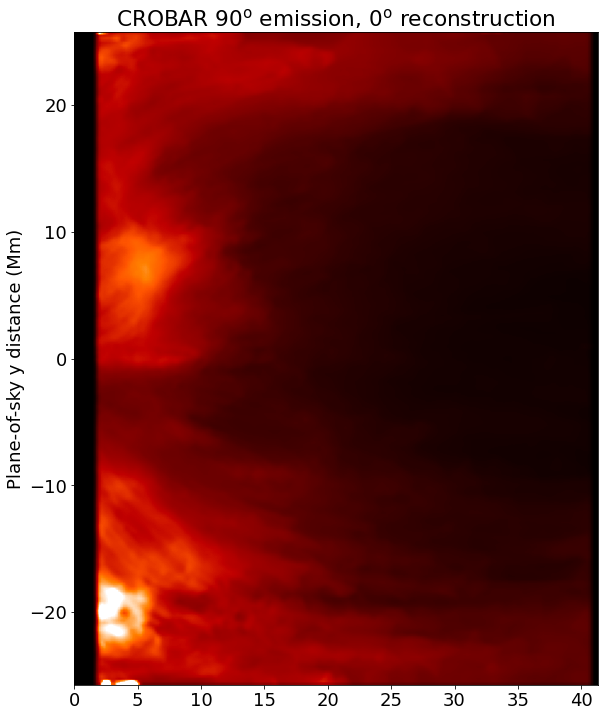}\\
	\includegraphics[width=0.5\textwidth]{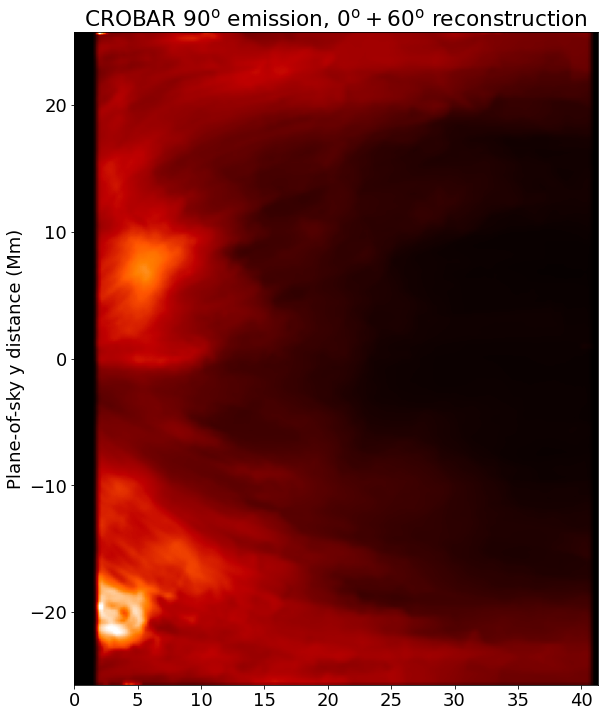}\includegraphics[width=0.5\textwidth]{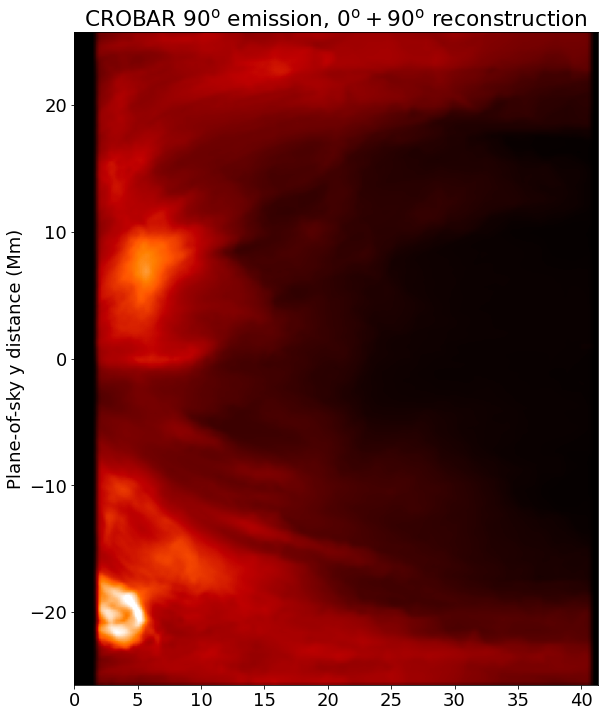}\\
	\caption{CROBAR reconstruction of the MURaM 90 degree emission. Top left shows original MURaM emission, top right shows CROBAR reconstruction using only the MURaM 0 degree emission, lower left shows CROBAR reconstruction using MURaM 0 and 90 degree emission, lower right shows CROBAR reconstruction using the MURaM 0 and 60 degree emission.}\label{fig:reconstructions_90}
 \end{figure}

\begin{figure}[!htbp]
	\includegraphics[width=0.5\textwidth]{MURaM_slice_90.png}\includegraphics[width=0.5\textwidth]{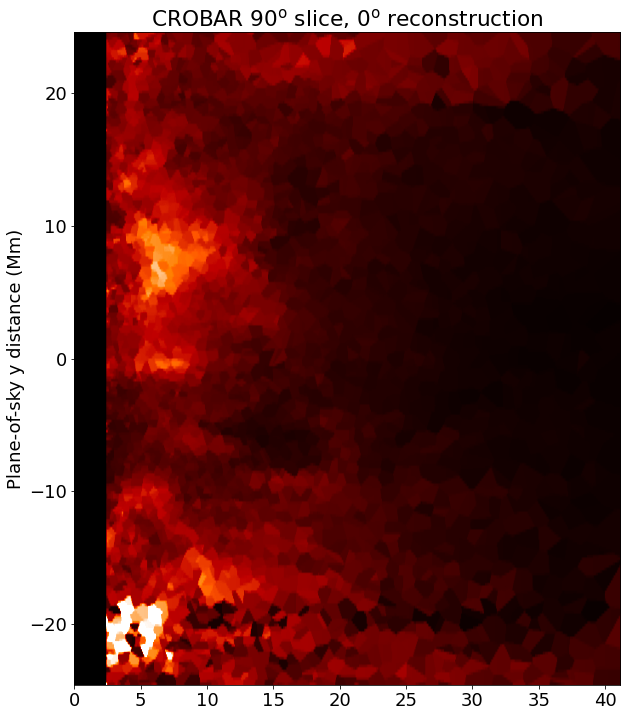}\\
	\includegraphics[width=0.5\textwidth]{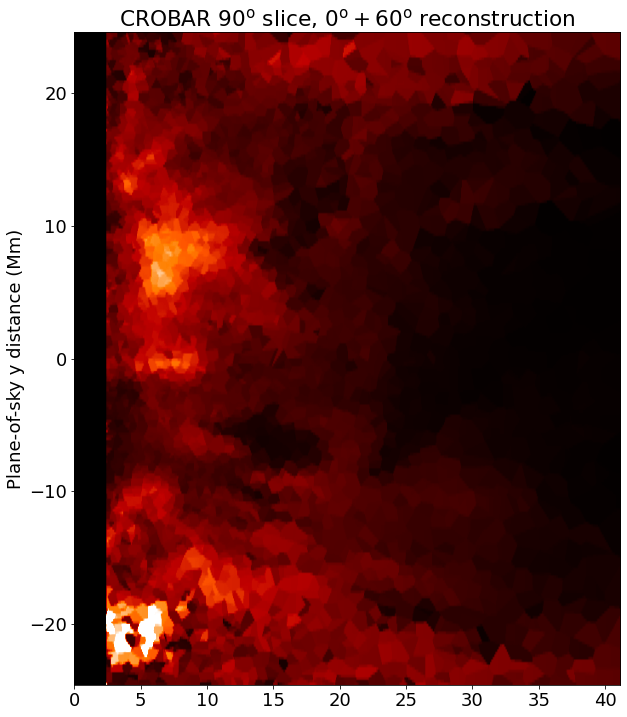}\includegraphics[width=0.5\textwidth]{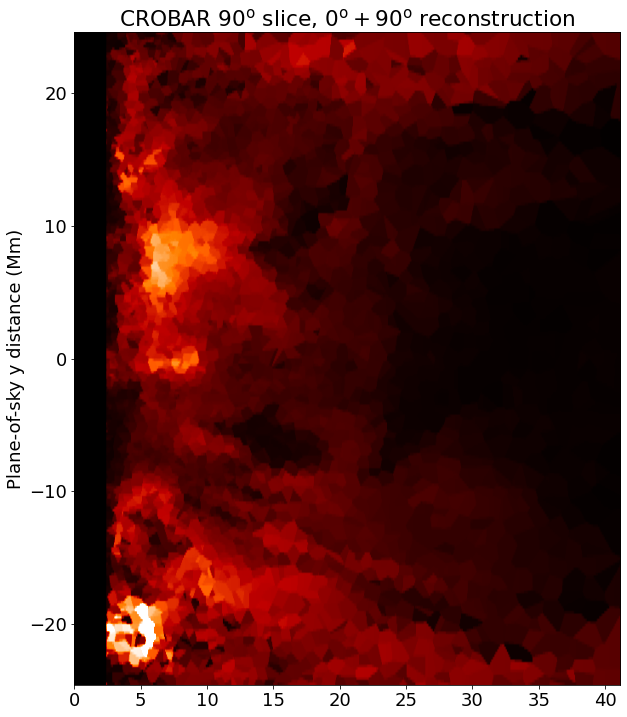}\\
	\caption{CROBAR reconstruction of a latitudinal slice the MURaM 0 degree emission. Top left shows original MURaM emission, top right shows CROBAR reconstruction using only the MURaM 0 degree emission, lower left shows CROBAR reconstruction using MURaM 0 and 90 degree emission, lower right shows CROBAR reconstruction using the MURaM 0 and 60 degree emission.}\label{fig:reconstructions_slice}
 \end{figure}

And now, the reconstruction results. I performed a single viewpoint reconstruction using only the overhead view, a two viewpoint reconstruction using the overhead and 60 degree views, and a two viewpoint reconstruction using the overhead and limb (90 degree) views. These multi-viewpoint reconstructions are easily done in CROBAR's framework by simply stacking the forward matrices, data vectors, and error vectors. Figure \ref{fig:reconstructions_overhead} shows the overhead view of all of these reconstructions, Figure \ref{fig:reconstructions_60} shows the 60 degree view, Figure \ref{fig:reconstructions_90} shows the 90 degree view, and lastly Figure \ref{fig:reconstructions_slice} shows the latitudinal slice also shown in figures \ref{fig:MURaM_fwd_synth_projections} and \ref{fig:BAR_example_plots}.

Overall, the results look quite good to my eyes. Even the single-perspective snapshot reproduces most of the structures seen in integrated emission, and many of them seen in the snapshot. This could likely be improved with an experimenting with the scale height of the field-aligned regions regions and allowing them to vary from one region to the next. The two perspective reconstructions are even better, with most structures in both the slice and the integrated emission being recovered. This includes the complicated 'veil'-like structures pointed out by \cite{MalanushenkoEtal_coronalveil2022}, which are best shown in the latitudinal slice (Figure \ref{fig:reconstructions_slice}. The overhead and 90 degree reconstruction reproduces the slice best, but this is no surprise considering it samples the vertical stratification best. To compared how well the different selection of perspectives constrains the volume emission as a whole, I performed an estimate of goodness-of-fit for the resulting cubes based on a median squared deviation using the following formula:

\begin{equation}
	\Delta E_{rel} \equiv \mathrm{Median}\Bigg[\Big(\frac{E_\mathrm{MURaM}(i,j,k)}{\mathrm{Median}[E_\mathrm{MURaM}]} - \frac{E_\mathrm{CROBAR}(i,j,k)}{\mathrm{Median}[E_\mathrm{CROBAR}]}\Big)^2\Bigg]
\end{equation}

\noindent ($E$ stands for emission and $(i,j,k)$ are the indices of the cubes -- for this expression, I bin down the CROBAR output so it matches that of the MURaM cube). By this metric, the overhead-only reconstruction scored 0.149, the 0 and 60 degree reconstruction scored 0.0549, and the 0 and 90 degree reconstruction scored 0.0589. A second perspective is therefore roughly three times as good by this metric than a single perspective, with 60 and 90 degrees being roughly comparable in their quality.

\section{Discussion, Implications, \& Conclusions}

I have demonstrated that CROBAR can reproduce the complex structures seen in the MURaM simulations reasonably well, provided a valid magnetic skeleton for its B-aligned regions. Even a single perspective can obtain useable results, and two perspectives are significantly better. This is despite some concerns \citep{MalanushenkoEtal_coronalveil2022} that the 'veil'-like nature of these structures (and presumably the real corona which MURaM attempts to model) might frustrate attempts to recover them using a limited number of perspectives. However, I would argue that despite their complex appearance in cross-section, we can expect these veil-like structures to still be governed by the field aligned (or nearly field aligned) condition: gradients of the plasma and emission along the field direction are small and can be represented by a small number of degrees of freedom through much of the volume. CROBAR's success in recovering this emission from the MURaM simulation is an indicator of the validity of this expectation. It also possesses the resolution and performance to resolve the complexity of the field-perpendicular structure. Being volume-filling is also an essential property of any decomposition which attempts to recover these structure, and CROBAR's field aligned regions possess this property as well.

\subsection{Implications for Coronal Missions Away from the Earth-Sun line}

Although CROBAR can provide 3D information from just a single perspective, the results shown here point out the importance of additional viewpoints and quantify the improvement from adding another viewpoint. They also indicate that both 60 and 90 degrees as additional perspectives provide similar improvements. This suggests that a 'drifter' spacecraft, moving from $\sim 45$ to 90 degrees over its lifetime, could provide a similar degree of reconstruction quality to one which was 'parked' (e.g., at a Lagrange point) over its entire lifetime. This would be similar to the orbits of the STEREO spacecraft. Orbits like that of Solar Orbiter, while usable for science  work, would be less desirable from a space weather standpoint due to inconsistent temporal coverage.

Most coronal space observations have focused on EUV lines or passbands with relatively low temperatures and narrow temperature coverage. However, some of the considerations discussed in Section \ref{sec:depth_related_considerations} indicate that these are not the best suited to the coronal 3D reconstruction problem. Instead, a passband with a temperature response that is a power law (or at least monotonic), peaked at high temperatures (hotter than the plasma being observed) is ideal. This suggests an X-ray imager (a la Yohkoh SXT or Hinode XRT) or perhaps a relatively broad-band EUV imager if a suitable wavelength window can be found. The other option would be to use DEMs, although that requires more lines/passbands (more complexity, weight, and telemetry) and must still include lines or passbands with high temperature response. That said, I have seen surprisingly promising results applying CROBAR to AIA passbands such as 335 \AA\ and 171 \AA , or the STEREO passbands (STEREO does not have much in the way of high temperature coverage, see comment below). I am still investigating this and it will be the topic of a future publication. Similarly for data from Solar Orbiter \cite{SolarOrbiter_AA2021}, particularly the SPICE spectrograph \citep{SPICEInstrument_AA2020}.

The short version of the above two paragraphs is that an `optimal' cost/benefit deep space mission to take advantage of CROBAR would include either an imager with soft-X-Ray-like temperature response functions (2 distinct channels) or an AIA like instrument with at least 5 channels and high temperature coverage at least as good as what's provided by the 335 \AA\ and 94 \AA\ channels (these are essential for AIA to cover the critical temperature range from $10^{6.4} to 10^{6.8}$ Kelvin, where many active region field lines have their temperature peaks). It should also observe from between 30 and 90 degrees. No spacecraft has done this to date; STEREO is the closest but its high temperature coverage is not well suited to CROBAR and the lifespan of the remaining spacecraft is highly uncertain.

\subsection{Conclusions and Next Steps}
The utility of CROBAR as a starting point, giving a three-dimensional emission structure corresponding to a coronal volume from just one or two vantage points, is substantial. It provides a viable path forward for future improvement: with it we can refine and optimize our models of the field with guidance from coronal emission observations, we can add refined models for the field aligned emission in accordance with modelers, and experiment with adding more detailed physics to the framework as opportunity and insight present themselves. Whereas without it we had been limited to speculating about how our 2D images connect to the actual 3D volume. 

This paper has focused on testing for the field-aligned reconstruction concept against the MURaM simulations, but I have already found significant success with initial implementation of the refinements just mentioned in the context of real solar data. I have already implemented a basic tunable field model, in the form of a nonlinear force-free-field, and applied it to CROBAR reconstructions with AIA and STEREO data with very promising results. These will be covered in an upcoming publication.

	\acknowledgements{The analysis in this work was carried out in Python, making use of NumPy \citep{numpy}, SciPy \citep{scipy}, AstroPy \citep{astropy, astropyII}, and SunPy \citep{sunpy}. I would like to acknowledge Matthias Rempel for providing the MURaM cube and Anna Malanushenko for some helpful discussion and the inital inspiration for CROBAR. This work was supported by NASA grant 80NSSC17K0598, by NASA under GSFC subcontract \# 80GSFC20C0053 to SwRI, and by SwRI internal research grant 15.R6225.
	}

\bibliographystyle{aasjournal}
\bibliography{CROBAR_MURaM_paper}

\end{document}